\documentclass[a4paper,12pt]{article}

\usepackage{amsfonts}
\usepackage{amssymb}
\usepackage{epsfig}

\def\beq{\begin{equation}}
\def\eeq{\end{equation}}
\def\bea{\begin{eqnarray}}
\def\eea{\end{eqnarray}}
\def\winf{W_{1+\infty}\ }
\def\u1{\widehat{U(1)}}
\def\su2{\widehat{SU(2)}_1}

\def\R{{\rm Re}\ }

\def\nl{\nonumber \\}

\catcode`\@=11
\begin{document}
\begin{titlepage}
\begin{center}
\hfill  \quad hep-th/0203228 \\ \vskip .5 in {\Large\bf A new
class of Matrix Models arising from the $\winf$ Algebra}

\vskip 0.5in
Henry David HERCE\\
{\em Department of Physics, North Carolina State University, Raleigh NC
27695, U.S.A.}
\\
\vskip 0.3in
Guillermo~Ra\'ul~ZEMBA
\footnote{
Fellow of CONICET, Argentina.}
\\
{\em Physics Department, Comisi\'on Nacional de Energ\'{\i}a
At\'omica, Av.Libertador 8250, (1429) Buenos Aires, Argentina and
Universidad Favaloro, Sol{\'{\i}}s 453, (1078) Buenos Aires,
Argentina}
\end{center}
\vskip .5 in
\begin{abstract}
We present a new class of hermitian one-matrix models originated 
in the $\winf$ algebra: more precisely, 
the polynomials defining the $\winf$ generators in their
fermionic bilinear form
are shown to expand the orthogonal basis of 
a class of random hermitian matrix models. The corresponding 
potentials are given, and the thermodynamic limit interpreted
in terms of a simple plasma picture.
The new matrix models can be successfully applied to the full
bosonization of interesting one-dimensional systems, 
including all the perturbative orders in 
the inverse size of the system.   
As a simple application,
we present the all-order bosonization of the free fermionic 
field on the
one-dimensional lattice.
\end{abstract}
\vfill \hfill March 2002
\end{titlepage}


The study of one-dimensional fermionic systems has recently
attracted much attention due to progress in manufacturing of
mesoscopic devices. Although much progress has been made in the
theoretical understanding of the dynamics of these 
strongly correlated systems, many
interesting questions remain still unanswered.
One theoretical tool that is successfully used is the 
{\it bosonization} of the fermionic degrees of freedom 
\cite{fermi}:
for a given fermionic system in the thermodynamic limit, 
one considers the low-lying particle-hole
excitations around the Fermi points. Linearization
of the dispersion relation around these
leads to a relativistic conformal 
field theory \cite{cft}
of these gapless effective degrees of freedom. More recently, an
extended bosonization procedure incorporating the subsequent
non-linear corrections of the actual dispersion relation has been
proposed \cite{clz,flsz}. The
{\it algebraic bosonization} technique is an all-order bosonization
approach based on the $\winf$ algebra, 
a characteristic fermionic symmetry \cite{shen}
extending the conformal 
symmetry of the standard bosonization. 
The main
advantage of this method is that it yields a systematic
expansion of the full hamiltonian (including interactions)
and all physical observables
in powers of the momentum fluctuations 
around the Fermi points, involving algebraic
procedures only.

In this letter, we present a one-parameter family of hermitian 
one-matrix models which originate in the $\winf$ algebra and 
are intimately connected to the algebraic bosonization 
of one-dimensional fermionic systems.
Consider the Fock space of a $(1+1)$-dimensional relativistic 
Weyl fermionic field with
creation and annihilation operators $a^{\dagger}_n$ and $a_n$,
respectively ($n$ is an integer that labels the momentum),
satisfying $\{a^{\dagger}_n,a_m\}=\delta_{n,m}$. 
We define the $\winf$ operators
$V^k_n$ ($k=0,1,2,\dots$) in this specific basis
\cite{kac,ctz1}: 
\beq 
V^k_n\ =\ 
\sum_{r\in {\bf Z}}\ P_n^k (r)\ {\bf :}\ a^{\dagger}_{r-n} a_r {\bf :}
\ ,
\label{defw}
\eeq 
which satisfy the $\winf$ algebra \cite{shen}, 
\beq
{[\ V^i_n, V^j_m\ ]} = (jn-im) V^{i+j-1}_{n+m}
+q(i,j,n,m)V^{i+j-3}_{n+m} +\cdots +\delta^{ij}\delta_{n+m,0}\ c\
d(i,n) \ , 
\label{walg} 
\eeq 
where the structure constants
$q(i,j,n,m)$ and $d(i,n)$ are polynomial in their arguments, $c$
is the central charge
($c=1$ for the single Weyl fermion), and the dots stand for a finite 
number of
terms involving the operators $V^{i+j-1-2k}_{n+m}\ $,
with $k=0,1,\dots,[(i+j-1)/2]$, where the brackets denote
the integer part function (the complete
expression of (\ref{walg}) is given in \cite{kac}). 
The meaning of
the operators $V^k_n$ is as follows: they parametrize
particle-hole excitations near the Fermi points, such that the
index $n$ accounts for the fixed momentum transfer, whereas the index
$k$ denotes the geometric moments (i.e.,``multipole moments'') of the 
deformation of the Fermi-Dirac sea \cite{ctz1,clz,flsz}. 
The explicit expression of the polynomials
$P_k^n(x)$ can be
inferred directly from the $\winf$ algebra \cite{odake}:


\beq 
P_n^k(x)\ =\ \frac{(-1)^k}{\left(\begin{array}{c} 2k\\k
\end{array}\right)
}\ \sum_{j=0}^{n}(-1)^j {\left(\begin{array}{c} n\\j
\end{array}\right)}^2
\left(n+1- x \right)_{k-j} \left( x \right)_{j}\ , 
\label{pfor}
\eeq
where $(a)_n = \prod_{\ell =0}^{n-1}
(a+\ell)$ denotes the Pochhammer symbol. Some properties can be
deduced from (\ref{pfor}): {\it i)} the polynomials have a
definite parity with respect to the reflection about $x=1/2$: 
even $n$ (resp.
odd) correspond to even (odd) polynomials. {\it ii)} for $n$
large enough, the zeroes of $P_n^k(x)$ are located on the line
$x=1/2 + i \zeta$, with $\zeta$ real, and for pure imaginary
values of $n$, all zeroes fall along this line.  For $k$ even, all
zeroes come in complex conjugate pairs, and for $k$ odd the same
is true except for a single zero at $x=1/2$. However, the most
important consequence of (\ref{pfor}) is that
the $P_n^k(x)$ satisfy a {\it three-term recurrence relation}:
\beq 
x P_n^k (x)\ =\ P_n^{k+1}(x)\ +\ \frac{1}{2}\ P_n^k(x)\ +\
\frac{k^2 ( n^2 - k^2)}{4(4k^2-1)}\ P_n^{k-1}(x)\ .
\label{recrel}
\eeq 
This property has the important consequence of
leading to a new class of random matrix models associated to the
$\winf$ algebra, as we shall show in the following.

Matrix models constitute another important tool when studying
systems displaying universal behavior in the thermodynamic limit 
\cite{matrix1}. There are many know examples in which the
symmetries that characterize the relevant effective degrees of
freedom of a given system are simply encoded in a specific
random matrix model,
which posses the advantage of being a simpler $(0+1)$-dimensional
theory.
Consider a generic random one-matrix model, defined by its partition 
function:
\beq 
Z(N)\ =\ \int d^{N^2}\ \Phi\ \exp\left[ - {\rm Tr} V(\Phi) \right]
\ , \label{zmm} 
\eeq 
where $\Phi$ is an $N \times N$ Hermitian
matrix, $\rm Tr$ is the trace, ane the real function 
$V(\Phi)$ is the potential. Eventually, we shall be interested
in the limit $N\to\infty$, which is the relevant thermodynamic
limit. 
The matrix $\Phi$ has real eigenvalues $x_i$,
$i=1,\dots,N$, in terms of which the partition function can be
written as \cite{matrix1}:
\beq 
Z(N)\ =\ \Omega(N)\ \int \dots \int \prod_{i=1}^N dx_i\
{\Delta _N}^{2}(x)\ \exp\left( - \sum_{i=1}^N V(x_i ) \right)\ ,
\label{zeig} 
\eeq 
where $\Delta_N (x) = \prod_{i<j} (x_j - x_i)$
and $\Omega(N)= (2\pi)^{N(N-1)/2}/\prod_{k=1}^N k!$ is 
the volume of the ``angular'' (non-diagonal) entries of
$\Phi$ \cite{matrix1}.
A general strategy to compute $Z(N)$ is to
factorize the $N$ integrals on the r.h.s. of (\ref{zeig}),
which can be done provided there exists a set of orthogonal
polynomials ${\cal P}_k(x)$ satisfying: 
\beq 
\int_{-\infty}^{\infty}\ dx\ {\rm e}^{-V(x)}
\ {\cal P}_k (x)\ {\cal P}_l (x)\ =\ h_k\ \delta_{k,l}\ .
\label{ort} 
\eeq 
In turn, this condition can be met 
provided the polynomials ${\cal P}_k
(x)$ satisfy a three-term recurrence relation given by: 
\beq 
x {\cal P}_k
(x)\ =\ {\cal P}_{k+1}(x)\ +\ S_k\ {\cal P}_k(x)\ +\ R_k\ {\cal P}_{k-1}(x)\ ,
\label{recrelma}
\eeq 
where $S_k$ and $R_k$ are specific to the matrix
model \cite{matrix1}. A crucial equality that relates the 
potential with
the recursion relation is $R_k = h_k/h_{k-1}$, and the partition
function is given by: 
\beq 
Z(N)\ = \Omega(N)\ N!\
\prod_{k=0}^{N-1} h_k\ =\ \Omega(N)\ N!\ h_0^N \prod_{k=0}^{N-1}\
R_k^{N-k}\ .
\label{zh}
\eeq

In the following, we shall focus in the set of polynomials
(\ref{pfor}), and show that they are orthogonal with 
respect to an specific measure which is then used to
define the associated matrix model.
One anticipates, therefore, an infinite family of models
parametrized by the momentum $n$, and defined by the 
family of potentials $V_n(x)$. Given the location of the zeroes of the
polynomials (\ref{pfor}), it is natural to perform a "Wick
rotation" in $x$ and define ${\cal P}_n^k(x) = i^k P_n^k\left( -ix
+ (n+1)/2 \right)$ as the set of $\winf$
orthogonal polynomials. 
They posses a definite parity under reflection in $x$;
the first few are:
${\cal P}_n^0(x)=1$, ${\cal P}_n^1(x)=x$,
 ${\cal P}_n^2(x)=x^2+(n^2-1)/12$,${\cal P}_n^3(x)=x^3+x(3n^2-7)/20$.
The recursion relation (\ref{recrel}) now takes the form: 
\beq 
x {\cal P}_n^k(x)\ =\ {\cal
P}_n^{k+1}(x)\ + \ \frac{k^2 ( k^2 - n^2)}{4(4k^2-1)} {\cal
P}_n^{k-1}(x)\ ,
\label{calrec}
\eeq 
which implies the existence of the set of potentials $V_n(x)$ 
making them orthogonal.
These functions are
readily found once one recognizes that the polynomials ${\cal
P}_n^k(x)$ are {\it Hahn polynomials}, for which the
orthogonality relations are known\cite{hahn}. 
There are, in fact, two classes of Hahn polynomials: discrete
and continuous. The first case arises for $n$ integer,
such that there is a finite set of polynomials, since
the recursion relation (\ref{calrec}) stops at $k=n$. 
However, given (\ref{defw}), we are interested in the case for 
which the family 
is infinite: this is the continuous case, that arises for
$n$ not an integer. For our purposes, it is convenient 
to consider the values $n=ip$, $p=0,1,2,\dots$ (i.e., 
imaginary momentum) since this yields a positive 
and non-degenerate inner product. This condition can be
viewed as the Wick rotation in momentum space that  
corresponds to the one previously performed in $x$.
Under these conditions, the orthogonality relations are given by
\cite{hahn}: 
\bea 
\int_{-\infty}^{\infty}\ dx\
&g_p(x)&\ {\cal P}_{ip}^k(x)\ {\cal P}_{ip}^l(x)\ = \frac{| \Gamma
(k+1+ip) |^2 }{(2k+1){\left(\begin{array}{c} 2k\\k
\end{array}\right)^2
}}\ \delta_{k,l}\ ,\nl
 &g_p(x)& = \frac{\pi}{ \cosh(2\pi x) +
\cosh (p\pi) }\ .
\label{ortint} 
\eea 
From this expression, one
extracts the family of potentials $V_p (x)$ 
defining the matrix models as in (\ref{zmm}):
\beq 
V_p (x)\ =\ -\ln  g_p(x)\ =\ \ln
\left [ \frac{\cosh(2\pi x)  + \cosh (p\pi)}{\pi} \right]\ 
\qquad , \qquad p=0,1,2,\dots\ ,
\label{potfam}
\eeq 
which asymptotically behave as $V_p (x) \simeq
2 \pi |x| - 2\ln (2\pi) $ for $|x| > \gamma_p/\pi$,
$\gamma_p = \ln [2\pi \cosh (p \pi)]$ (see figure 1). 
\begin{figure}
\epsfysize=7cm
\centerline{\epsfbox{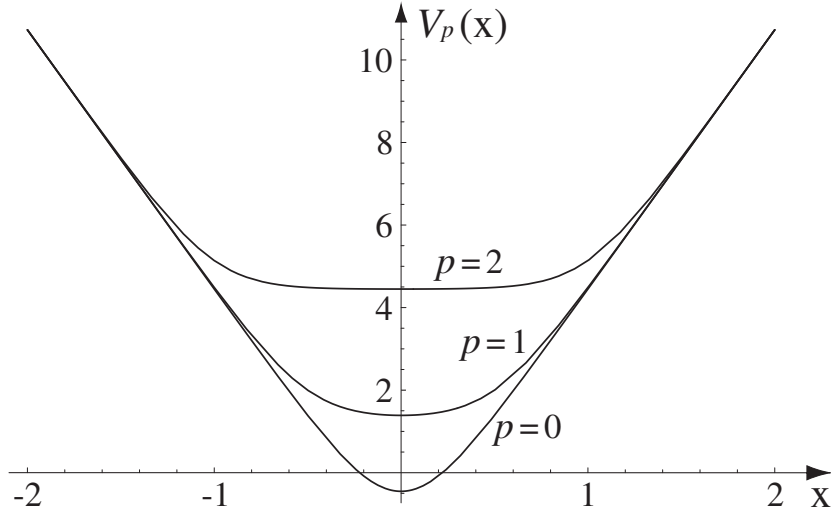}}
\caption{The potentials $V_p(x)$ as a function of $x$ for $p=0,1,2$.
Note the asymptotic linear behavior discussed in the text.}
\end{figure}
From (\ref{ortint}), one also knows that
the norm of the $k-th$ polynomial in the $p-th$ matrix model is: 
\beq 
h_p^k\ =\ \frac{| \Gamma
(k+1+ip) |^2 }{(2k+1){\left(\begin{array}{c} 2k\\k
\end{array}\right)^2
}}\ ,
\label{hnorm}
\eeq 
which is in agreement with the recursion relation
(\ref{calrec}) through the equality $R_p^k = 
h_p^k/h_p^{k-1}= k^2(k^2 + p^2)/4(4k^2 -1)$.
From (\ref{hnorm}), one can therefore define the partition function
$Z_p(N)$ and free energy density $f_p(N)$
for the $p-th$ model as:
\beq 
Z_p(N)\ = \Omega(N)\ N!\
\prod_{k=0}^{N-1} h_p^k\ =\ \exp(N^2 f_p(N))\ .
\label{zmod}
\eeq
The asymptotic form of  $f_p(N)$ 
in the thermodynamic limit $N\to\infty$
is:
\beq
f_p(N)\ =\ \frac{1}{2}\ln N\ -\ \left( \frac{3}{4} +
\frac{1}{2}\ln \frac{8}{\pi} \right)\  +\ O\left(\ln N/N\right)\ .,
\eeq
independent of $p$ to the given order. 

The thermodynamic limit leads also to a simple
physical picture in terms of a fictitious one-dimensional 
plasma: consider the partition function (\ref{zmm}) for the
potential (\ref{potfam}) (we consider for simplicity the
case $p=0$). The asymptotic form of $V_0(x)$ is linear, and
suggests the definition of a new scaling variable $y=x/N$,
appropriate in the limit $N\to\infty$. This implies that 
$V_0(x) \simeq 2 \pi N|y|-2\ln(2\pi)$ for $|y| > 1/\Lambda$, 
with $\Lambda=\pi N/\gamma_0$. In the 
scaling limit, $y$ is defined in the entire real axis,
except for a small region of the size of the cutoff $1/\Lambda$
around the origin. 
We can write the partition function approximately as: 
\beq
Z_0(N)\ \simeq\ \frac{\Omega(N)}{N^N}\ \int \prod_{i=1}^N dy_i\
\exp\left[ - N\left(2\pi\sum_{i=1}^N |y_i | 
- \frac{1}{N}\sum_{i<j} \ln (y_i - y_j)^2 + f(N)
\right)\right]\ ,
\label{asymz} 
\eeq
where $f(N)=(N-1)\ln N + 2\ln(2\pi)$.
In the large $N$ limit, $Z_0(N)$ is dominated by the 
saddle-point configurations $\{ y_1,\dots,y_N \}$ 
satisfying the equations:
\beq
2 \pi\  {\rm sg}(y_i)\ =\ \frac{2}{N} \sum_{j, j \neq i}\ \frac{1}
{y_i - y_j}\ ,\qquad i=1,\dots,N\ ,
\label{speqd}
\eeq
where ${\rm sg}(y)$ is the sign function. The scaled variable
$y$ is appropriate for 
taking the continuum limit: $y_i \to y$, $\sum_i (\dots)/N \to
\int dy \rho(y) (\dots)$, where $\rho(y)$ is the {\it density function}.
The saddle-point equations are:
\bea
\pi\  {\rm sg}(y)\ &=& {\cal P}\int_{-\infty}^{+\infty}
\frac{\rho(y')}{y - y'}\ dy' \ , \\
\int_{-\infty}^{\infty} \rho(y')\  dy ' &=& 1\ ,
\label{speqc}
\eea
where ${\cal P}$ denotes the Cauchy principal part of the 
integral, and the second equality is the normalization condition.
These equations give a semiclassical physical interpretation
to our system: they describe a system of fictitious 
equally charged
``particles'', located on a one-dimensional rod, and 
subject to a neutralizing background. The particles
repel each other with a two-dimensional Coulomb force,
while they are attracted with constant force
to an infinite two-dimensional
uniformly charged plane, that is perpendicular to the rod.
One can find a solution to equations (\ref{speqc}) by
assuming that the system is symmetric under $y \to -y $,
and therefore the total number of ``particles'' is even,
with $N/2$ of them on each side of the infinite charged plane.
Therefore, one seeks for a solution of the form 
$\rho(y ) = \rho_+ (y) E(y) + \rho_- (y) E(-y)$, with
$E(y)$ the Heaviside function. In the following we shall focus 
on one side only, e.g. $(+)$. As it is customary, one
introduces the function
\beq 
F_+ (y)\ =\ \int_{-\infty}^{\infty}
\frac{\rho(y')}{y - y'}\  dy' \ , 
\label{f+def}
\eeq
defined on the complex half-plane $\R (y) > 0$. 
Assuming that this function has one cut running 
along the real axis from $0$ to  
$\alpha$, one can solve (\ref{speqc}) by a 
standard procedure \cite{matrix1}, yielding: 
\beq
F_+ (y)\ =\ \pi\ - \pi\ \sqrt{\frac{y-\alpha}{y}}\ ,
\label{f+}
\eeq
From (\ref{f+}) one determines
the density:
\beq
\rho_+ (y)\ = \ \sqrt{\frac{\alpha-y}{y}}\ , 
\label{rho}
\eeq
where $\alpha=1/\pi$ is determined from the normalization
condition (\ref{speqc}).
One can verify that along the cut and close
to its extreme $y \simeq \alpha$, the behavior of 
$ \rho_+ (y)$
is in agreement with the expected one for a linear
background potential \cite{matrix1}. 

Let us now focus on the application of the 
previous results to one-dimensional condensed matter models.
Consider a free non-relativistic fermionic field on a 
one-dimensional lattice of unit spacing. 
We shall outline the effective field theory formulation of this problem
in terms of the $\winf$ symmetry generators \cite{flsz}.
While this description does not appear as necessary in the free
theory, it becomes rather advantageous when including 
the effects of the interactions, as, for example, in the $XXZ$
Heisenberg model. We shall 
postpone the full discussion of the effects of the interactions
for a forthcoming publication, but remark here that the scope of
our approach is to develop a bosonization scheme valid to all
orders in the expansion parameter, which is the inverse size of the
system, given by $N$.
In the thermodynamic limit, the non-relativistic field can be 
expanded in the basis of the relativistic Weyl field, and the Fermi
sea interpreted as a Dirac sea \cite{flsz}.
Moreover, the free hamiltonian ${\cal H}$ 
can be written as the sum of two independent 
terms, ${\cal H}_+ $ and ${\cal H}_- $,
describing the fluctuations around the
right $(+)$ and left $(-)$ Fermi points 
at momenta $\pm n_F$, respectively, with
$n_F = (N-2)/4$; i.e., we consider
the ground state of the system to be at half-filling, 
for simplicity.  
Both terms are decoupled, up to global (zero-mode)
constraints. Therefore,
we shall consider $ {\cal H}_+$ only in the following:
\beq
{\cal H}_+\ =\ -\ \sum_{r=-\infty}^{+\infty}\ 
f(n_F + r)\ {\bf :} a^{\dagger}_r a_r {\bf :}\ \qquad ,
\qquad
f(n) = \cos \left( \frac{2\pi}{N} n\right)\ ,
\label{freeh}
\eeq
where $a_r$ are the fermionic Fock operators 
of the Weyl field, obtained from the original ones by
shifting the momentum index,
and $r$ is integer. 
The dispersion relation around the Fermi point 
given by $ f(n_F + r)$ is non-linear, and can
be expanded in terms of the zero-momentum $\winf$ polynomials 
(\ref{pfor}) as 
\cite{flsz}:
\beq
{\cal H}_+\ =\ \sum_{k=0}^{\infty}\ c_k \ V^k_0\ ,
\eeq
where the coefficients $c_k$ are such that:
\beq
\sin \left[ \frac{2\pi}{N} \left( r -\frac{1}{2} \right)
\right]\ =\ \sum_{k=0}^{\infty}\ c_k\ P^k_0 (r) \ .
\eeq
Making the change of variables $r \to ( -i r + 1/2)$ and
passing to the basis ${\cal P}^k_0 (r)$ allows 
us to determine $c_k$ using the orthogonality relation
(\ref{ort}):
\beq
c_k\ = \frac{i^k}{h_0^k}\ \int_{-\infty}^{+\infty}
\ dx\ g_0(x)\ \sin \left( \frac{2\pi}{N} r \right)\ 
{\cal P}^k_0 (x)\ \qquad k=0,1,2,\dots\ ,
\label{overlap}
\eeq
where for convenience we have made the redefinition
$2\pi /N \to i 2 \pi /N$, so as to deal with regular 
expressions; this analytic continuation should be undoed
at the end of the computations.
From (\ref{overlap})
one immediately concludes that $c_{2n}=0$ for $n$ integer,
and defines $C_k = c_{2k+1}$ ($k=0,1,2,\dots$) for the other cases, 
which can be worked out by considering the set 
of integrals:
\beq
I_n(y)\ =\ \int_{-\infty}^{+\infty}
\ dx\ g_0(x)\ {\rm e}^{i x y}\  {\cal P}^n_0 (x)\ ,
\ n=0,1,2,\dots\ .
\label{in}
\eeq
As a consequence of (\ref{calrec}), they satisfy the 
recursion relations:
\beq
I_{n+1}(y)\ =\ -i \frac{d I_n (y)}{d y}\ - R^n_0\ I_{n-1}(y)\ 
\qquad ,\qquad
I_0 (y)\ =\ \frac{y}{2 \sinh (y/2) }\ ,
\eeq
which are formally solved by 
$I_n(y)\ =\ {\cal P}^n_0 \left( -i \frac{d}{d y} \right) I_0 (y)$.
The coefficients (\ref{overlap}) are then readily determined
yielding:
\beq
C_k\ =\ \frac{(-1)^k}{h^{2k+1}_0}\ I_{2k+1}
\left( \frac{2\pi}{N} \right )\ ,\  k=0,1,2,\dots.
\eeq
As and example, after analytically continuing back $\pi/N \to -i \pi/N$
as explained above, 
\beq
C_1\ =\ \frac{6}{\sin (\pi/N)}\left[ 1\ -\ \frac{\pi}{N\tan (\pi/N)}
\right]\ =\ \frac{2\pi}{N}\ +\ \frac{7}{120}\left(\frac{2\pi}{N}
\right)^3\ +\ O\left\{\left(\frac{2\pi}{N}\right)^5\right\}\ ,
\eeq
which reproduces the results of \cite{flsz}. We have also verified 
the leading term in $C_2= -(2\pi/N)^3/6 + \dots$ yields the correct result.

Finally, we remark that the matrix models
considered in this letter belong to the class of models with
{\it non-polynomial potentials}, which include also the 
Penner \cite{penner}
and Frustrated Spherical \cite{qpol} models.
According to the `t Hooft correspondence, one can
associate to each matrix model a tiling of a 
two-dimensional random  surface, such that the 
the number of sides and the ways of gluing the elementary poligons 
are determined by the algebraic powers of the monomials that define
the potential \cite{thooft}\cite{matrix1}.
The class of matrix models with non-polynomial potentials 
possess the characteristic property of having an unbounded
set of polygons and gluing vertices among them.
We conclude by remarking that it would be interesting to
explore whether other infinite-dimensional Lie algebras 
ecould give rise to further new classes of random matrix models. 
 
GZ acknowledges the hospitality of the
Abdus Salam Center for Theoretical Physics (Italy) and the Physics
Department and I.N.F.N. of Florence (Italy).

%
\def\NP{{\it Nucl. Phys.\ }}
\def\PRL{{\it Phys. Rev. Lett.\ }}
\def\PL{{\it Phys. Lett.\ }}
\def\PR{{\it Phys. Rev.\ }}
\def\CMP{{\it Comm. Math. Phys.\ }}
\def\IJMP{{\it Int. J. Mod. Phys.\ }}
\def\MPL{{\it Mod. Phys. Lett.\ }}
\def\RMP{{\it Rev. Mod. Phys.\ }}
\def\AP{{\it Ann. Phys.\ }}

\end{document}